\documentclass{article}
\usepackage{spconf,amsmath,graphicx}

\usepackage{xcolor}
\usepackage{float}
\usepackage{dblfloatfix}

\title{Forward Error Correction applied to JPEG-XS codestreams}
\name{Antoine Legrand, Benoît Macq, Christophe De Vleeschouwer
}
\address{ICTEAM, UCLouvain, Belgium
}
\begin{document}
\ninept
\maketitle
\begin{abstract}
JPEG-XS offers low complexity image compression for applications with constrained but reasonable bit-rate, and low latency. Our paper explores the deployment of JPEG-XS on lossy packet networks. To preserve low latency, Forward Error Correction (FEC) is envisioned as the protection mechanism of interest. Although the JPEG-XS codestream is not scalable in essence, we observe that the loss of a codestream fraction impacts the decoded image quality differently, depending on whether this codestream fraction corresponds to codestream headers, to coefficient significance information, or to low/high frequency data. Hence, we propose a rate-distortion optimal unequal error protection scheme that adapts the redundancy level of Reed-Solomon codes according to the rate of channel losses and the type of information protected by the code. Our experiments demonstrate that, at 5\% loss rates, it reduces the Mean Squared Error by up to 92\% and 65\%, compared to a transmission without and with optimal but equal protection, respectively. 
\end{abstract}
\section{Introduction}

    
    JPEG-XS has become the \textit{de facto} standard in computationally efficient and low-latency compression of high-resolution images. It is now widely used since applications that require the transmission of high resolution visual content under low-latency constraints have largely gained in interest \cite{interactive_video_streaming}.
    Despite this success, JPEG-XS does not satisfy the needs for image transmission created by the emergence of applications such as wireless TVs or wireless virtual reality headsets. 
    This limitation arises from the fact that (i) 
    wireless links are prone to errors, which significantly degrade the transmitted image quality in the absence of error protection, and (ii) wireless links exhibit lower bandwidths than their wired counterpart, thereby requiring a rate-effective protection, so as to preserve a sufficient fraction of the channel rate for source coding.


    To facilitate JPEG-XS deployment in wireless environments, our work first observes that a partial loss of a JPEG-XS codestream impacts the decoded image quality quite differently depending on whether the lost bytes are related to the codestream headers, the coefficient bitplanes semantic, or the low/high frequency data. As a second contribution, our study then leverages this image decoding quality discrepancy to design a rate-distortion optimal Unequal Error Protection (UEP) scheme, which jointly selects the source coding rate and the variable level of protection of the different parts of a JPEG-XS codestream to minimize the expected decoded image distortion for a given channel, whose bit- and loss-rates are known.
    
    Our experiments demonstrate that the proposed transmission scheme rapidly outperforms unprotected transmission when losses are present (up to 6dB average PSNR gains at 5 \% loss rate), but also improves a rate-distortion optimal equal protection of the JPEG-XS codestream (above 1dB average PSNR gains at 5 \% loss rate when the channel rate becomes lower than 400kB/frame).
    
    
    The paper is organized as follows. Section 2 surveys the main trends in previous works related to protection of image streams on packet-switched and wireless channels. Section 3 introduces the structure of a JPEG-XS codestream, and analyzes how its decoding is impacted by the channel losses. Section 4 presents our proposed UEP framework, and derives the equations to select the rate-distortion optimal source coding and codestream protection parameters, for given channel bandwidth and loss features. Section 5 presents the benefits of our UEP solution compared to unprotected and equally protected transmission, while Section 6 concludes.

\section{Related Work}
    
     Image and video communication on best effort packet-switched networks is challenging. Solutions include priority allocation of network resources to images \cite{receiver_driven_TCP}, fine adjustment of compression rate to the channel capacity \cite{adaptive_streaming, videoconf, model_based_rate_control}, error concealment mechanisms \cite{error_concealment, SSIM_error_resilient}, and use of scalable codecs to adapt the codestream, on-the-fly, to the channel at hand \cite{jpeg2000_coding_system, scalable_epitomes, remote_interactive, prefetching}.
     
    In the presence of packet losses, Forward Error Correction (FEC) codes are the preferred solution in case of strong latency requirements. FEC adds parity bits to the informative ones, so that the decoder can recover the initial information as long as a sufficient fraction of transmitted bits are correctly received~\cite{RS}. The burstiness of losses, as typically induced by the channel packetization, is known to hamper the correction capabilities of FEC codes. Therefore, data are generally interleaved or transmitted on multiple-path\cite{path_diversity}. Another important practical aspect related to the deployment of FEC lies in the selection of the level of redundancy associated with the correction code.
     A uniform FEC protection implicitly assumes that all the transmitted bytes are equally important/informative regarding the image decoding. On image codestreams, this assumption does not hold \cite{bytes_not_equal}. Hence, many papers have considered the use of unequal protection to protect classes of different importance within a compressed image/video stream \cite{ULP_framework_assignment, JP2K_Quality_layers_RS_Baruffa, UEP_Bitplanes, JointSourceChannel_JPEG2000_1, JointSourceChannel_JPEG2000_2, UEP_JP2K_Natu, JP2K_Headers_RS}. Those solutions are specific to the codec at hand and exploit its specificities, i.e. progressive decoding capabilities\cite{ULP_framework_assignment}, quality layers\cite{JP2K_Quality_layers_RS_Baruffa}, or bitplanes\cite{UEP_Bitplanes}. To the best of our knowledge, none of them has been designed to handle JPEG-XS, probably because JPEG-XS does not naturally offer scalability.
    
     
\section{JPEG-XS in front of losses}
   
        A JPEG-XS codestream consists of groups of consecutive lines, named precincts, that are independently encoded. Each precinct is made of Low-Frequency (LF) and High-Frequency (HF) packets, each packet being split into significance, bitplane count, and data subpackets\cite{ISO_JPEG_XS,XS_latency}. The data subpackets encode the wavelet coefficients associated with the group of lines as a set of bitplanes. To improve entropy coding efficiency, significance and bitplane count subpackets provide information on which coefficients are significant and the number of bitplanes used to represent those coefficients, respectively.\footnote{To further improve entropy coding efficiency, the JPEG-XS standard provides a feature known as Sign Packing\cite{ISO_JPEG_XS}. However, as this feature increases the intra-stream dependencies, it significantly decreases its ability to cope with corrupted data. Hence, in the rest of this paper, we only consider images compressed without Sign Packing.}
        Headers are added to the datastream, to each precinct and to each packet, to facilitate decoder resynchronization in case of errors. They are required to recover any corrupted codestreams.
        
\begin{figure}[t]
            \centering
            \includegraphics[trim = 15mm 0mm 15mm 5mm, clip, ,scale=0.45]{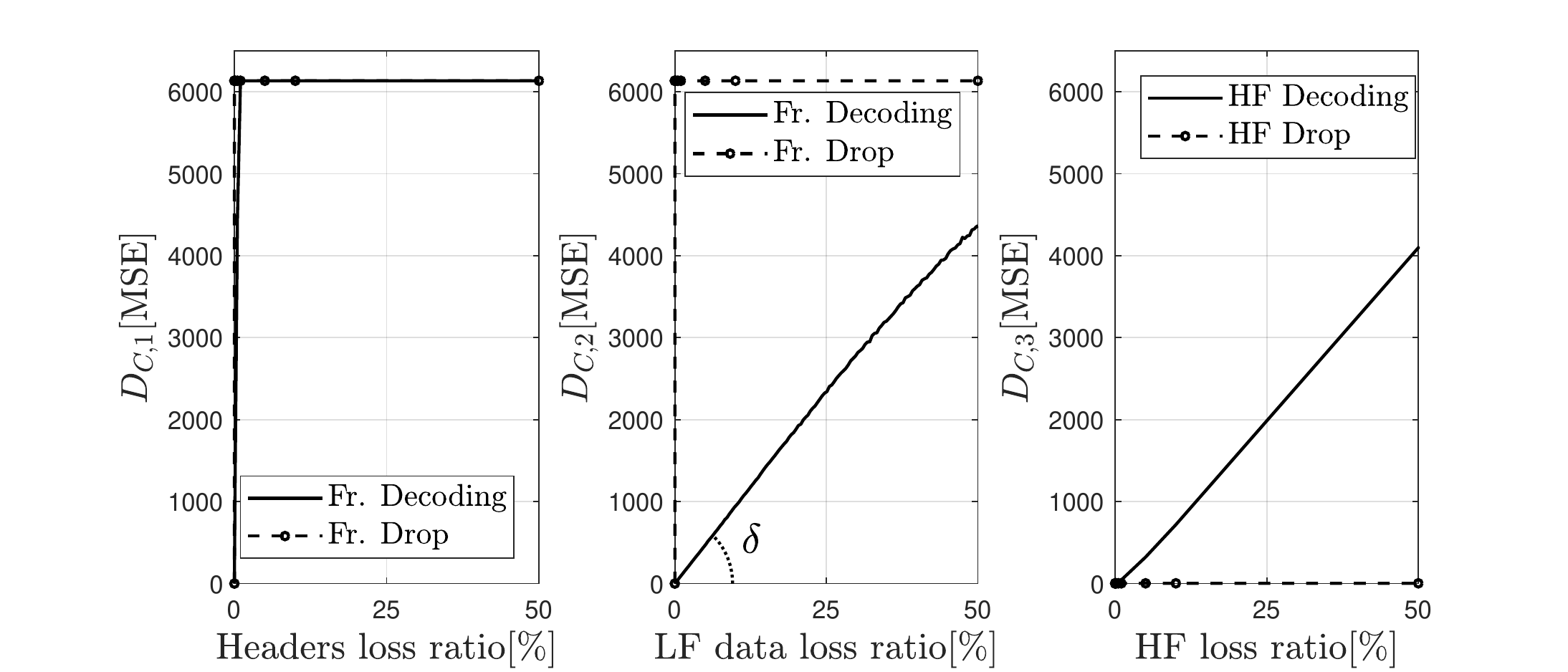} 
            \caption{Distortion induced by the corruption of the first (left), second (middle) and third (right) classes, averaged on 100 error patterns applied to the \textsc{Beauty} frame, and decoded either with or without dropping the whole corrupted class. Unlike class 2, the best decoding strategy regarding classes 1 and 3 consists in not decoding them when they are corrupted.}
            \label{fig_XS_Dist}
        \end{figure}

        Since the different types of subpackets in a codestream have very different purposes, one might expect that their loss impacts differently the decoded image quality. Hence, it becomes relevant to design an Unequal Error Protection framework that better protects most impactful subpackets.. 
        Therefore, three distinct classes of information are considered. They are defined in decreasing order of importance as follows:
        
        \textbf{Class 1.} {\it Headers}, {\it Low-Frequency significance and bitplane count} subpackets. The loss of any header byte induces the loss of the whole frame. Similarly, losing {\it Low-Frequency significance or bitplane count} bytes severely degrades the decoded image, since it induces the loss of wavelet coefficients on a large spatial area. 
        Hence, the distortion associated with the loss of even a small part of those bytes is set equal to the loss of the whole frame in our rate-distortion optimization. The legitimacy of this strategy is confirmed by the plot presented on the left of Figure \ref{fig_XS_Dist}. It depicts how the decoded image distortion increases with the loss of an increased fraction of bytes of the first class, assuming other classes are perfectly received. We observe the distortion rapidly increases to the one obtained when dropping the whole frame.
        
        \textbf{Class 2.} {\it Low-Frequency data} subpackets. In a JPEG-XS codestream, the distortion caused by the loss of low-frequency data bytes remains limited to the corresponding wavelet coefficients.  This behavior is depicted in the middle part of Figure \ref{fig_XS_Dist}, where we observe that the distortion induced by the loss of this class is proportional to the number of lost bytes, while remaining largely below the distortion associated with full frame dropping.. Hence, losing parts of this class should not cause the rejection of the full frame. Instead, decoding a distorted image from the corrupted stream generally results in a smaller distortion than simply ignoring the codestream (and decoding a grey image). 
        
        \textbf{Class 3.} {\it High-Frequency content} subpackets. The High-Frequency content aims at enhancing the image sharpness and details. When only a fraction of HF packets is received, their decoding results in corrupted HF wavelet coefficients, leading to significant noise to the image. This phenomenon is reflected by the right plot of Figure \ref{fig_XS_Dist} which reveals that omitting the entire High Frequency subpackets when some of them are corrupted induces a much lower distortion than wrongly interpreting them. Hence, in the following, when this part of the bitstream is corrupted, it is ignored at decoding.
        
\begin{figure}[t]
            \centering
            \includegraphics[trim = 7mm 0mm 13mm 4mm, clip, ,scale=0.555]{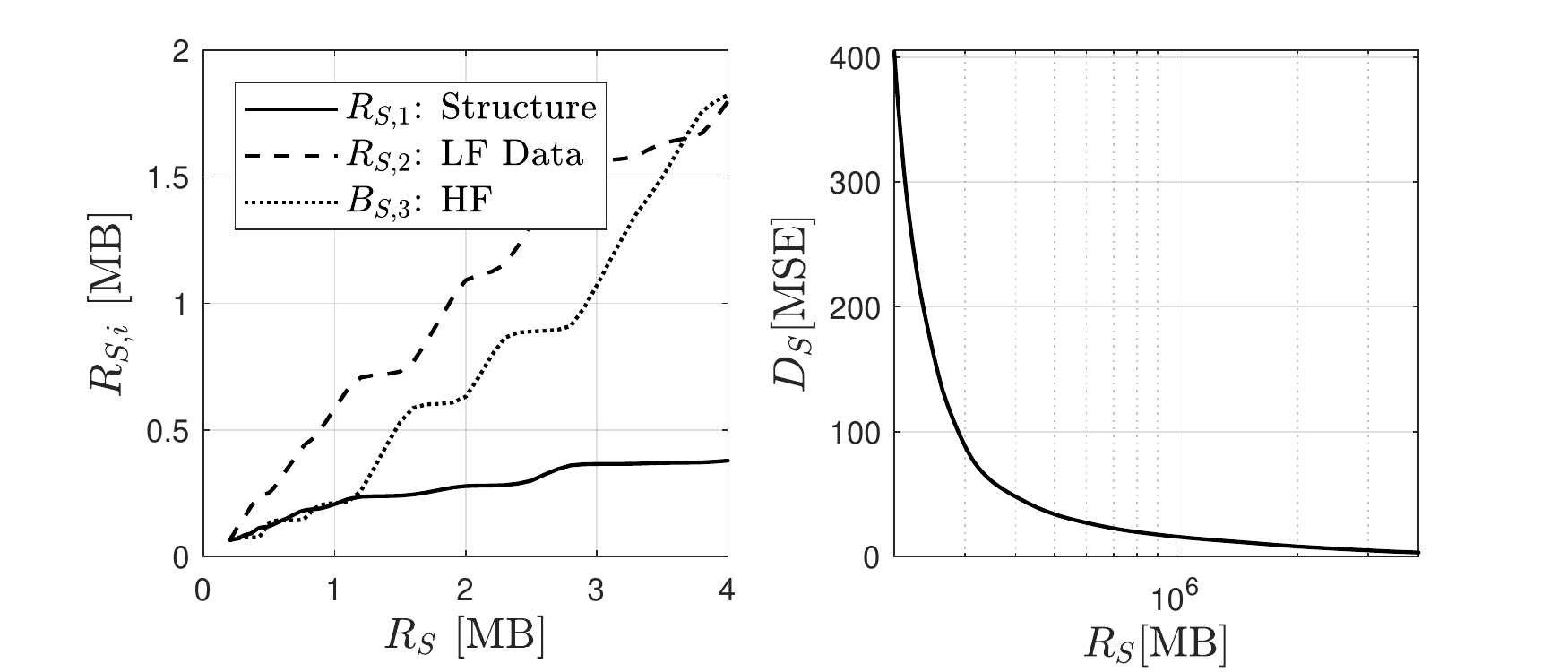} 
            \caption{\textbf{(left)} JPEG-XS Rate allocation among the different classes. \textbf{(right)} JPEG-XS Rate-Distortion curve. Both graphs depict the values averaged on 12 frames of the UVG dataset\cite{UVG_dataset}}
            \label{fig_RSiRS}
            \label{fig_rate_dist}
        \end{figure}

        The relationships between the encoding rate of JPEG-XS and the size of the different classes is depicted in Figure \ref{fig_RSiRS} (left).

\section{UEP FOR JPEG-XS}
\label{sec:UEP}

This section introduces our proposed unequal error protection (UEP) scheme. It first presents the codestream interleaving mechanism (Section 4.1), and then develops a rate-distortion source and code rates allocation method (Section 4.2). Eventually, it comments about the Equal Error Protection (EEP) case (Section 4.3). 

\subsection{Interleaving JPEG-XS}
\label{subsec:interleave}
    
\begin{figure}[t]
        \centering
        \includegraphics[trim = 0mm 0mm 0mm 0mm, clip, ,scale=0.35]{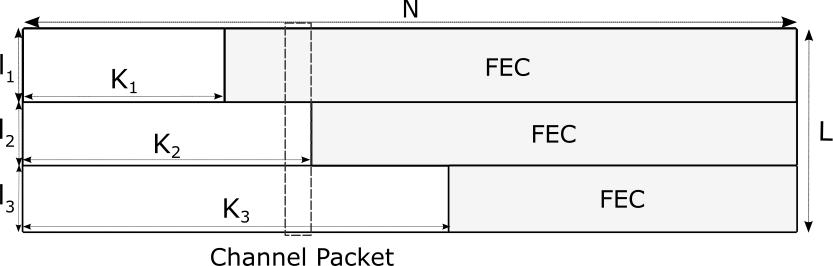}
        \caption{Interleaving block of our Unequal Error Protection Framework. Each of the 3 classes is encoded into $l_i$ Reed-Solomon codewords over GF($2^8$)\cite{RS, RS_open_source}. Hence, each codeword is made of N bytes (N$<2^8$) of which $K_i$ bytes come from the JPEG-XS codestream, the rest being parity bytes.The interleaving block is thus made of N network channel packets. As long as at most N-$K_i$ packets are lost, the $i^{th}$ class can be entirely recovered.}
        \label{fig_UEP_Framework}
    \end{figure}

    The UEP structure of interleaving blocks adopted to implement UEP and transmit the three classes of a JPEG-XS codestream on a lossy packet-switched network is depicted in Figure \ref{fig_UEP_Framework}. Due to the finite channel packet length (typically 1500 bytes) and the limited number of bytes within a Reed-Solomon code over GF($2^8$) (N$\leq$255), the codestream, whose size reaches about 1MB, is split in several blocks that are independently encoded according to Figure \ref{fig_UEP_Framework}.  It is worth noting that a precinct is never spread over multiples blocks. Hence, each interleaving block may be considered as independent from the other ones.
    
\subsection{Optimal selection of the transmission parameters}
    \label{sec_method}
    
    This section proposes a method to select the compression and protection parameters so as to maximize the expected reconstruction quality (i.e. minimize the distortion) when transmitting JPEG-XS codestreams on a known lossy packet switched network. The terminology adopted in the rest of the paper is summarized in Table \ref{tab:Model_Terminology}.
        
    \begin{table}[t]
        \begin{center}
            \caption{\label{tab:Model_Terminology} Terminology}           
            \begin{tabular}{ |p{1.35cm}||p{6.15cm}|}
                \hline
                \footnotesize{Y} & \small Random variable denoting the number of lost channel packets per interleaving block.\\
                \hline
                \footnotesize{\textbf{K}} & \small Vector denoting denoting the number of information symbols used by the Reed-Solomon codewords encoding each class: \textbf{K}=$\left[K_1,K_2,K_3\right]^T $.\\
                \footnotesize{\textbf{K}}/N & \small FEC code rate vector.\\
                \hline
                \footnotesize{$R_{S} $} & \small Source coding bit-budget in bytes.  \\
                \footnotesize{$R_{S,i}$} & \small Size of the $i^{th}$ class in a codestream compressed at $R_S$ (see left plot of Figure \ref{fig_RSiRS}). \\
                \footnotesize{$R_{C}$} & \small Channel coding bit-budget in bytes: \hspace{1cm} {\footnotesize$R_C = N \sum\limits_{i=1}^{3} \frac{R_{S,i}}{K_i}$}.\\
                \hline
                \footnotesize{$D_{T}(R_S, \textbf{K})$} & \small Total distortion caused by the transmission. \\
                \footnotesize{$D_S(R_S)$} & \small Distortion due to the compression (see right plot of Figure \ref{fig_rate_dist}).\\
                \footnotesize{$D_C$(\textbf{K})} & \small Distortion induced by the channel losses. \\
                \footnotesize{$D_{C,i}(K_i)$} & \small Distortion resulting from the corruption of the $i^{th}$ class by the channel losses. \\
                \hline
                \footnotesize{$\lambda$} & \small Lagrange multiplier to the Rate-Distortion optimization problem, i.e. $\lambda=\frac{\partial D_{T}}{\partial R_{C}}$.\\
                \hline
            \end{tabular}
        \end{center}
    \end{table} 
   
    The distortions induced by the compression and by the losses on the channel are assumed to be independent. Hence, the total distortion  $D_T(R_S,\textbf{K})$ expected when transmitting a JPEG-XS image codestream of size $R_S$, protected with a code rate vector \textbf{K}/N, is written:
    \begin{equation}
        \footnotesize D_{T}(R_S,\textbf{K}) = D_{S}(R_S) + D_{C}(\textbf{K})
    \end{equation}
    with $D_S(R_S)$ denoting the distortion induced by JPEG-XS source coding, and $D_C(\textbf{K})$ referring to the distortion caused by channel losses. In this notation, the channel parameters are omitted, and only the parameters that are subject to optimization (source rate $R_S$, and number of information bytes per codeword $K_i$) are mentioned.

    Rate-distortion optimality implies that a change of channel rate impacts similarly the distortion, whether this change of rate impacts the source coding rate or the channel coding. Formally, we have:    
    \begin{equation}
        \footnotesize \lambda = \frac{\partial D_{S}}{\partial R_{C}}  =  \frac{\partial D_{C}}{\partial R_{C}}
    \end{equation}
    
    \subsubsection*{Source Distortion}
    By applying the chain rule, 
    \begin{equation}
       \footnotesize \frac{\partial D_{S}}{\partial R_{C}} = \frac{\partial D_{S}}{\partial R_{S}} \frac{\partial R_{S}}{\partial R_{C}}
        \label{eq_Source}
    \end{equation}
    where $\frac{\partial D_{S}}{\partial R_{S}}$ is obtained from the Rate-Distortion curve depicted in Figure \ref{fig_rate_dist}(right), and $\frac{\partial R_S}{\partial R_C}$ is approximated by $\frac{R_S}{R_C}$, thereby assuming that the change in $R_S$ is distributed among the classes proportionally to the fraction of $R_S$ assigned to those classes. As can be observed in Figure \ref{fig_RSiRS}(left), this hypothesis is reasonably valid, especially at moderate rates, which are the ones of practical interest.
    
    \vspace{-0.3cm}
    \subsubsection*{Channel Distortion}
    \vspace{-0.1cm}
    
    Under the assumption that the different classes may be considered as independent, $D_C$ may be approximated by:
    \begin{equation}
        \footnotesize D_{C} = \sum_{i=1}^{3} D_{C,i}.
    \end{equation}
    Given the orthogonality between Low-Frequency and High-Frequency coefficients in a wavelet-based encoder and the significantly higher distortion associated with the loss of the first class compared to the corruption of the other ones, this independence assumption may be considered as relevant.
    As a consequence, optimality requires:
    \begin{equation}
        \footnotesize \lambda = \frac{\partial D_{C,i}}{\partial R_{C,i}} \quad \forall \quad i=1,2,3
    \end{equation}
    \begin{equation}
        \footnotesize \lambda =\frac{\partial D_{C,i}}{\partial K_i} \frac{\partial K_i}{\partial R_{C,i}} = \frac{-K_i^2}{N R_{S,i}} \frac{\partial D_{C,i}}{\partial K_i} \quad \forall \quad i=1,2,3
        \label{eq_Channel}
    \end{equation}
    
    For i=1 and 3, we assume that the image (for i=1) or the high-frequency class (for i=3) is not decoded at all when the number of packets lost by the channel exceeds the correction capabilities of the FEC. This is because the corruption of information associated with those classes rapidly degrades the image quality up to (or even above) the level where this information is ignored (see Figure 1, left and right).  Hence,
    \begin{equation}
       \footnotesize D_{C,i} = \Delta_i p(Y>N-K_i) \quad \forall \quad i=(1,3)
       \label{eq_DC13}
    \end{equation}
    where $\Delta_1 = \Delta_{ALL}$ denotes the distortion induced by the loss of the whole image and  $\Delta_3 = \Delta_{HF}$ denotes the distortion induced by the loss of the High Frequency content.
    
    In contrast, for the second class, we observe in the middle plot of Figure \ref{fig_XS_Dist} that the expected distortion associated with the loss of a fraction of the class information increases proportionally to the fraction of lost information. Letting $\delta$ denote the rate of distortion increase per percent of lost packets, we have:
    \begin{equation}
       \footnotesize D_{C,2} = \sum_{j=N-K_2+1}^{N} 100 \frac{j}{N} \delta \quad p(Y=j) + 0 \quad p(Y \leq N-K_2) 
       \label{eq_DC2}
    \end{equation}
    
    Given Eq. (\ref{eq_DC13}) and (\ref{eq_DC2}), $\frac{\partial D_{C,i}}{\partial K_i}$ can be approximated from the partial derivatives of p(Y) with respect to $K_i$, which directly depend on the known channel model.
    
    Overall,  for a given lambda, $R_S$ and \textbf{K} are derived from equations \ref{eq_Source} and \ref{eq_Channel}. This system of 4 equations is solved iteratively. Each iteration consists in two steps. The first step considers that \textbf{K} is fixed (with initialization to $K_i$ = N, for all i), and updates $R_S$. The second one updates \textbf{K}, for a fixed $R_S$. 

    \subsection{EEP case}
    
    In the case where $K_i = K$ for all i, the interleaving block is either correctly decoded or ignored, when the number of lost channel packets exceeds the FEC correction capability. 
    Hence, for a fixed channel rate $R_C$, the optimal EEP parameter (K) is computed so as to minimize the following expression:
    \begin{equation}\footnotesize D_T = D_S\left(R_C \frac{K}{N} \right) + \Delta_{ALL} \quad p(Y>N-K),
    \end{equation}
    with the same definition of variables as above.

\section{Results}
    \label{sec_results}
    
    \begin{figure}[t]
        \centering
        \includegraphics[trim = 7mm 7mm 16mm 7mm, clip, ,scale=0.555]{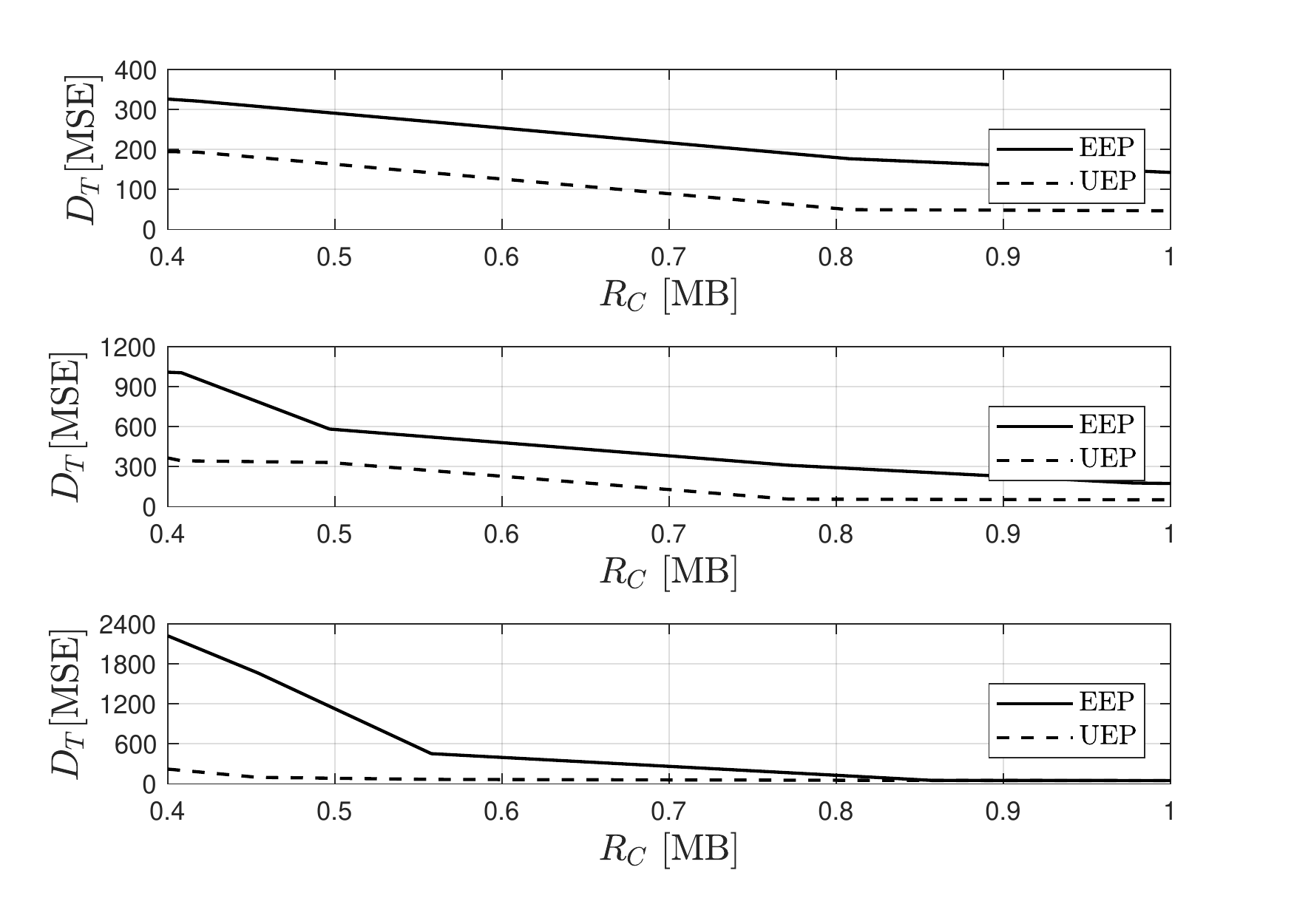}
        \caption{Mean Squared Error of the transmission of the \textsc{Beauty} frame at packet loss rates of: \textbf{(top)} 1\%, \textbf{(middle)} 5\% and \textbf{(bottom)} 10\%. For any channel and loss rate, UEP outperforms EEP, in particular at low channel rates and high packet loss rates}
        \label{fig_UEP_main}
    \end{figure}

    This section compares UEP and EEP, and demonstrates their benefit compared to an unprotected transmission. It first presents the considered simulation settings. Then, the performance comparison is discussed in terms of Mean Squared Error, average PSNR and ratio of decoded frames.

    \vspace{-0.3cm}
    \subsubsection*{Simulation settings}    
    \vspace{-0.2cm}
    
    Our experiments are conducted on a same UHD image, from the UVG dataset\cite{UVG_dataset}, corrupted by 500 error-patterns generated by a simple Gilbert model\cite{Gilbert_Model} whose parameters are computed to provide an average burst error length of 20 packets at a given packet loss rate. Other channel settings are considered in appendix \ref{appendix_results}.
    
    The transmission parameters are optimized according to the method described in Section~\ref{sec:UEP}, using $\delta$=90, $\Delta_{ALL}$=9000 and $\Delta_{HF}$=4. Those parameters correspond to the means of the corresponding metrics computed over the first frame of 12 sequences of the UVG dataset\cite{UVG_dataset}. The interleaving blocks are composed of N=255 packets of L=1500 bytes, so as to maximize the Reed-Solomon codes efficiency and match the characteristics of real networks, respectively. For typical JPEG-XS codestream sizes, such an interleaving block size induces a latency that is less than a frame (typically 40\% of a frame for a 1MB codestream).
    
    
    \vspace{-0.3cm}
    \subsubsection*{Mean Squared Error}
    \vspace{-0.2cm}
    
    Figure \ref{fig_UEP_main} depicts the MSE associated with the transmission of the \textsc{Beauty}\cite{UVG_dataset} frame on channels exhibiting 1, 5 or 10\% of packet losses, when protected with our UEP and EEP schemes. An unprotected transmission leads to MSE ranging around 1000, 5000, 10000 at packet loss rates of 1, 5, and 10\%, respectively (not depicted, to limit the MSE range in the plots). Hence, our proposed protection schemes are effective, since they decrease the MSE by a factor typically ranging from 5 to 10 (e.g. 92\% MSE decrease at 400kB/frame and 5\% loss rates). Interestingly, for any packet loss rate and any channel rate $R_C$, the UEP framework also outperforms EEP. This is especially true at low channel rates and large loss rates. On a 400kB/frame channel, with 5\% of losses, the UEP MSE is reduced by 65\%, compared to EEP. 
    
    Similar improvements have been obtained on other frames of the UVG dataset. Numerically, around 400kB/frame, the mean gains in MSE of UEP over EEP typically reach 40\%, 65\%, and 90\% at 1, 5, and 10\% channel loss rates, respectively.

    \begin{figure}[t]
        \centering
        \includegraphics[trim = 9mm 6mm 2mm 7mm, clip, ,scale=0.5]{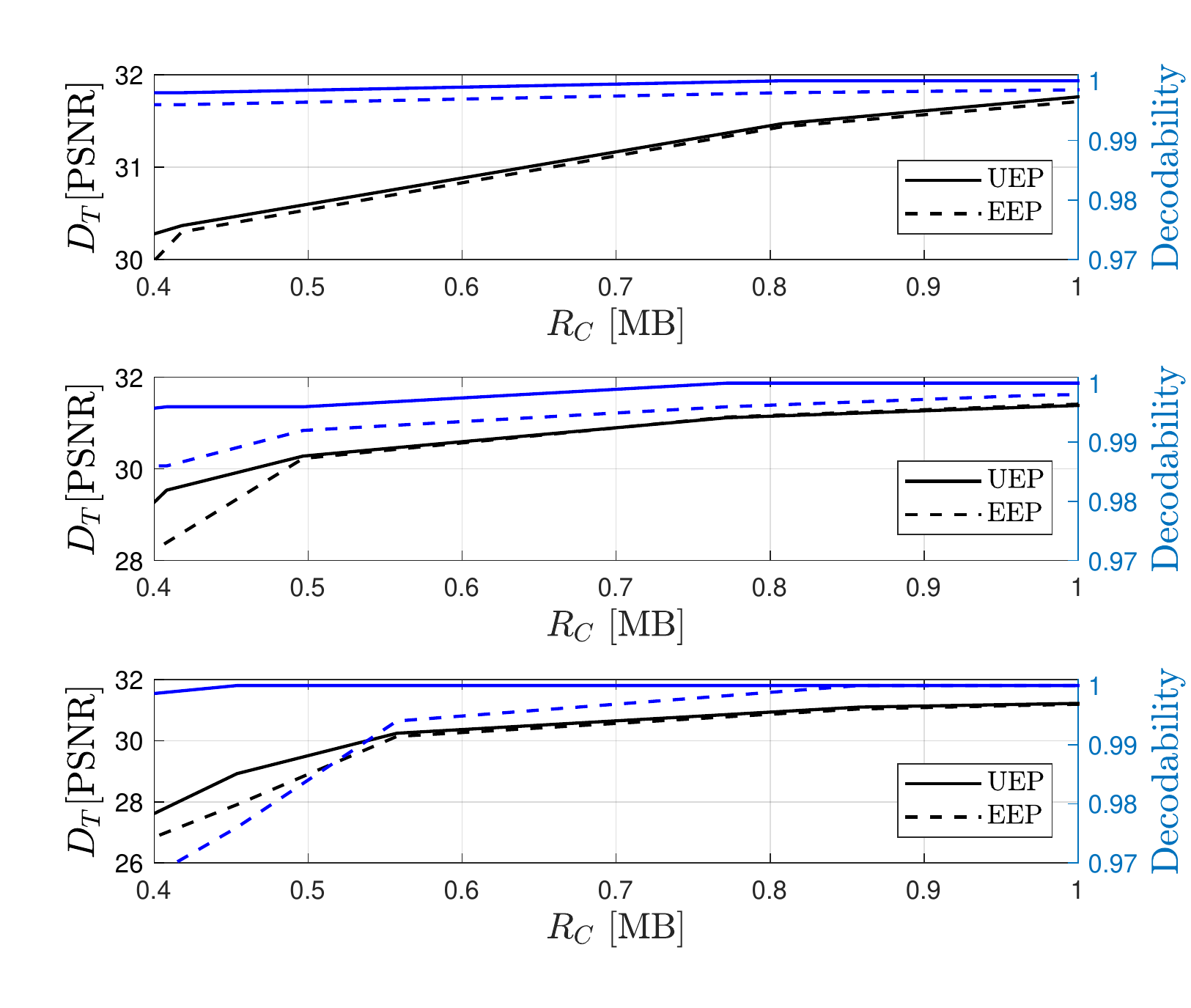}
        \caption{Average PSNR of the decoded \textsc{Beauty} and ratio of decoded frames when transmitted at packet loss rates of: \textbf{(top)} 1\%, \textbf{(middle)} 5\% and \textbf{(bottom)} 10\%. For any channel and loss rate, UEP outperforms EEP on both metrics, in particular at low channel rates and high packet loss rates}
        \label{fig_UEP_main_PSNR_Avg_received}
    \end{figure}
    
    \vspace{-0.3cm}
    \subsubsection*{Average PSNR and ratio of decodable frames}
    \vspace{-0.2cm}
    
    Figure \ref{fig_UEP_main_PSNR_Avg_received} analyzes the benefit of UEP over EEP in terms of average PSNR and ratio of decoded frames. As a preliminary remark, it is worth mentioning that both protection schemes outperform an unprotected transmission which reaches about 29, 23, and 17dB at 1, 5, and 10\% loss rates, respectively. Hence, at a channel rate of 400MB/frame and a loss rate of 5 \%, unequally protecting the codestream enhances by more than 6dB the average PSNR compared to to unprotected transmission. In Figure \ref{fig_UEP_main_PSNR_Avg_received}, we observe that UEP improves EEP both in terms of decoded frame ratio and average decoded frames PSNR quality, reducing the fraction of lost frames by a factor of 3 (from about 1.5\% to less than 0.5\%) , and, reaching gains up to 1dB at 400kB/frame at 5\% channel loss rate.
    


\section{Conclusions}

Our work has investigated the transmission of JPEG-XS on lossy packet networks. Therefore, it has proposed to partition the bytes of a JPEG-XS codestream depending on how their loss impact the decoded image quality. Three classes of bytes have been defined. The first class includes headers, low-frequency wavelet coefficient significance and bitplane counts, while the second and third correspond to low-frequency data and to high-frequency information (significance + data). In the presence of corruption, our study reveals that only the second class deserves to be partly decoded. Other classes should better be ignored when partly corrupted. This observation has been exploited to design rate-distortion optimal equal or unequal protection schemes. Our experiments demonstrate significant benefits (up to 6 dB at 5\% channel loss rates) compared to unprotected transmission. They also reveal that unequal error protection allows to convey a larger ratio of frames at higher PSNR values than its equal protection counterpart.

    
    
    


\newpage

\bibliographystyle{IEEEbib}
\bibliography{strings,refs}
\newpage

\begin{figure*}[!hb]

    \appendix
    \begin{minipage}[l]{0.48\linewidth}
        \section{Additional results}\label{appendix_results}

        \textbf{Performance on channels with different loss patterns}
        
        This section discusses the ability of our scheme to handle different channel properties. Figures \ref{fig_Add_PSNR_Bernoulli}, \ref{fig_Add_PSNR_ABEL_10}, \ref{fig_Add_PSNR_ABEL_20} and \ref{fig_Add_PSNR_ABEL_30} depict the average PSNR and ratio of decoded frames as a function of the channel rate, for different loss patterns. Figure \ref{fig_Add_PSNR_Bernoulli} considers channels whose losses follow Bernoulli distributions while Figures \ref{fig_Add_PSNR_ABEL_10}, \ref{fig_Add_PSNR_ABEL_20} and \ref{fig_Add_PSNR_ABEL_30} consider channels whose losses follow a Gilbert model with average burst error lengths of 10, 20 and 30 packets, respectively. Apart from the channel settings, all simulations parameters are kept from Section \ref{sec_results}.
        
        \vspace{0.5cm}

    \end{minipage}
    \hfill
    \begin{minipage}[l]{0.48\linewidth}
        In all our experiments, the Unequal Error Protection framework outperforms or is on par with the optimal Equal Error Protection allocation. In particular, at low channel rates, our Unequal Error Protection significantly outperforms an optimal Equal Error Protection scheme on channels with large packet loss rates occurring in bursts. 
        \vspace{0.5cm}
    \end{minipage}

    \begin{minipage}[c]{0.485\linewidth}
        \includegraphics[trim = 9mm 6mm 2mm 7mm, clip, ,scale=0.5]{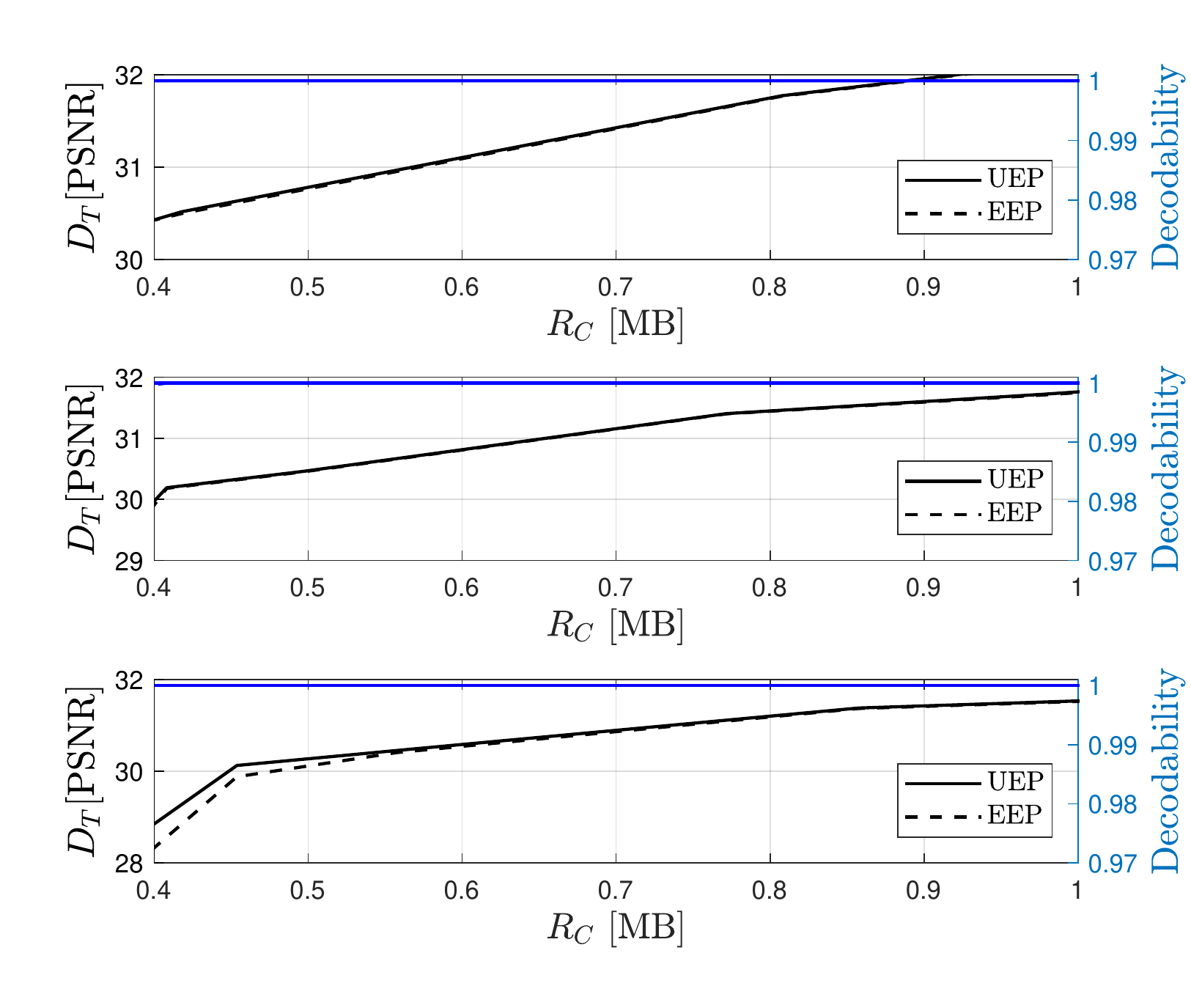}
        \caption{Average PSNR of the decoded \textsc{Beauty} and ratio of decoded frames when transmitted at packet loss rates of: \textbf{(top)} 1\%, \textbf{(middle)} 5\% and \textbf{(bottom)} 10\%, on a channel whose losses follow a Bernoulli distribution.}
        \label{fig_Add_PSNR_Bernoulli}
    \end{minipage}
    \hfill
    \begin{minipage}[c]{0.485\linewidth}
        \includegraphics[trim = 9mm 6mm 2mm 7mm, clip, ,scale=0.5]{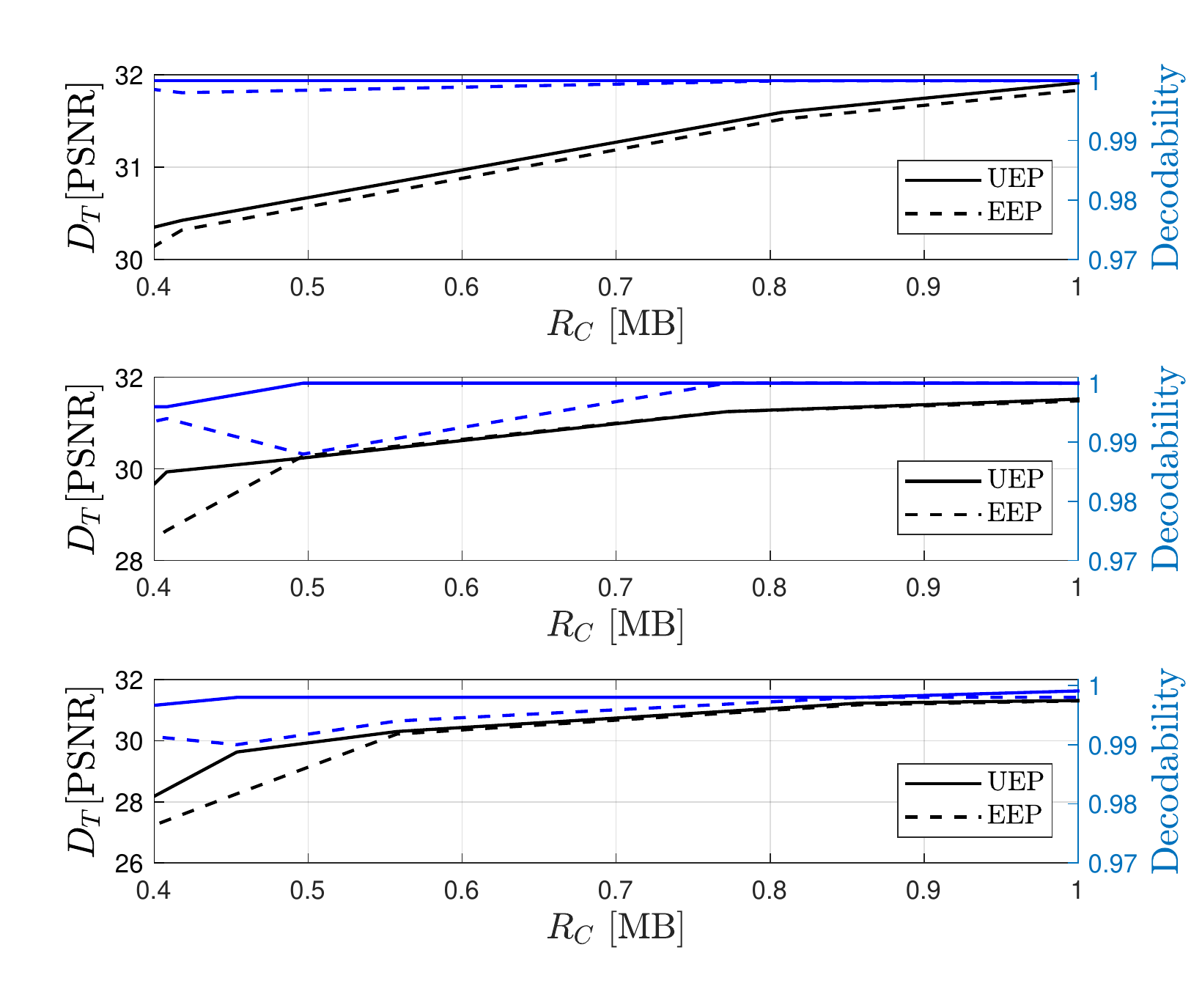}
        \caption{Average PSNR of the decoded \textsc{Beauty} and ratio of decoded frames when transmitted at packet loss rates of: \textbf{(top)} 1\%, \textbf{(middle)} 5\% and \textbf{(bottom)} 10\%, on a channel whose losses follow a Gilbert model with an Average Burst Error Length of 10 packets.}
        \label{fig_Add_PSNR_ABEL_10}
    \end{minipage}\\
    
    \begin{minipage}[c]{0.485\linewidth}
        \includegraphics[trim = 9mm 6mm 2mm 7mm, clip, ,scale=0.5]{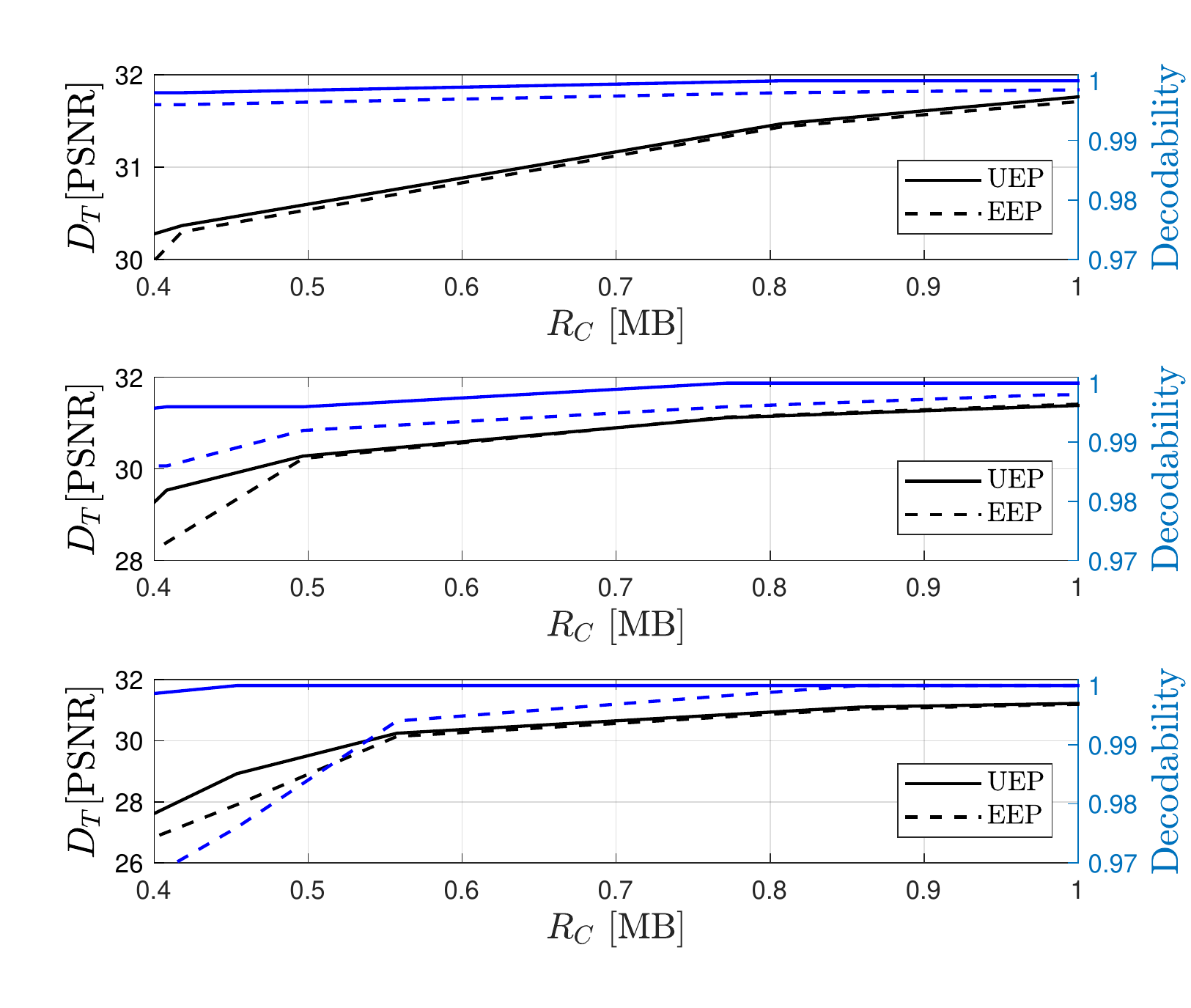}
        \caption{Average PSNR of the decoded \textsc{Beauty} and ratio of decoded frames when transmitted at packet loss rates of: \textbf{(top)} 1\%, \textbf{(middle)} 5\% and \textbf{(bottom)} 10\%, on a channel whose losses follow a Gilbert model with an Average Burst Error Length of 20 packets.}
        \label{fig_Add_PSNR_ABEL_20}
    \end{minipage}
    \hfill
    \begin{minipage}[c]{0.485\linewidth}
        \includegraphics[trim = 9mm 6mm 2mm 7mm, clip, ,scale=0.5]{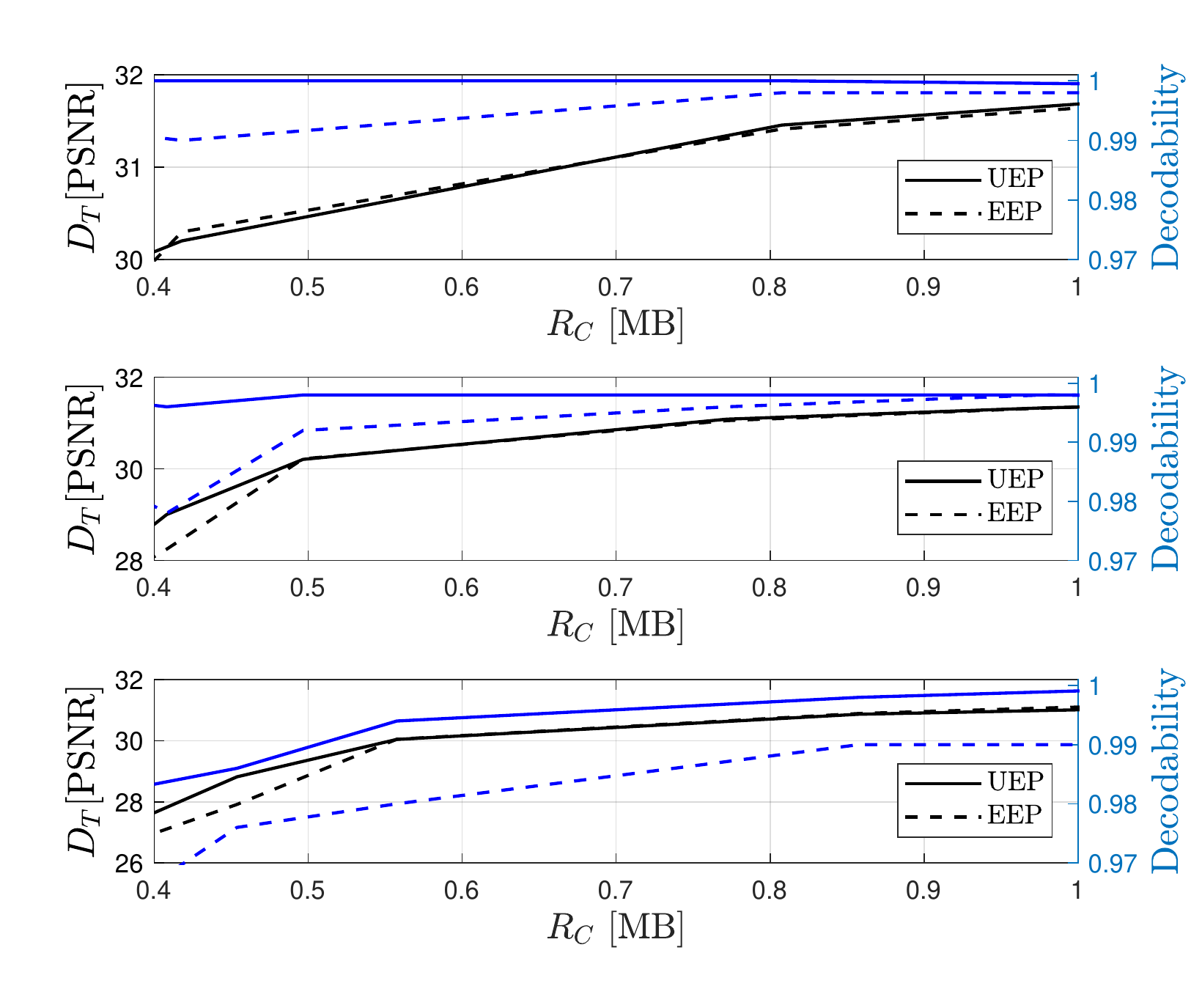}
        \caption{Average PSNR of the decoded \textsc{Beauty} and ratio of decoded frames when transmitted at packet loss rates of: \textbf{(top)} 1\%, \textbf{(middle)} 5\% and \textbf{(bottom)} 10\%, on a channel whose losses follow a Gilbert model with an Average Burst Error Length of 30 packets.}
        \label{fig_Add_PSNR_ABEL_30}
    \end{minipage}
        
\end{figure*}

\end{document}